# Renormalons*

C.T. Sachrajda[a]

[a]Dep. of Physics, University of Southampton,
Southampton SO17 1BJ, U.K.

The calculation of higher twist (or dimension) corrections to physical quantities using operator product expansions is delicate. If dimensional regularization is used to regulate the ultra-violet divergences then there are ambiguities in the Wilson coefficient functions due to infra-red renormalon singularities. With a hard ultra violet cut-off, such as the inverse lattice spacing $a$, there are no renormalon ambiguities, as a result of cancellations between terms which in finite orders of perturbation theory diverge as inverse powers of $a$, and those which diverge at most logarithmically. In this lecture I review these questions, explaining the steps necessary to obtain predictions for physical quantities from lattice measurements of matrix elements of higher dimensional operators. The ideas are illustrated by considering quantities computed using the heavy quark effective theory beyond leading order in the heavy quark mass.

## 1. Introduction

In recent months there has been renewed interest in the evaluation of power corrections to QCD predictions for hard scattering processes and related quantites, and in particular in attempting to understand the implications of the presence of renormalon singularities. Among the important physical quantities which are being studied are the higher-twist contributions to deep inelastic structure functions, the $1/m_Q$ corrections to masses and operator matrix elements in heavy quark physics (where $m_Q$ is the mass of the heavy quark $Q$), and corrections to predictions for event shape variables in jet physics and to the cross-section for the Drell-Yan process (i.e. the production of lepton pairs with a large invariant mass in hadronic collisions). For some of these processes the discussion can be formulated in terms of the operator product expansion, and it is such processes which will be considered in detail in this talk. The relevance of renormalons in field theory, and for operator product expansions in particular, has been developed in the papers listed in ref.[1]. This work has been exploited and ex-tended to the study of renormalons in the Heavy Quark Effective Theory (HQET) in refs.[2]-[6]. There has however also been some work recently on the rôle of renormalons in hard scattering processes for which there is no operator product expansion (see refs [7]-[12] for example). In this talk I will discuss the implications of the existence of renormalons in general, and in evaluating physical quantities from lattice simulations in particular. This may seem a surprising subject for a lecture, since renormalons are a consequence of the divergent behaviour of perturbation theory, and lattice simulations are a non-perturbative technique. It should be remembered however that in determining physical quantites from operator matrix elements computed on a lattice, one has to go through the process of "matching" which is performed in perturbation theory. It is at this stage that the question of the existence and importance of renormalons arises. The study of renormalons in lattice field theory, with its "hard" ultaviolet cut-off $a^{-1}$ (where $a$ is the lattice spacing), also provides important general insights, complementing the usual studies which are almost exclusively carried out using dimensional regularization of ultra-violet divergences.

In lattice field theory higher dimensional oper-

---

ators mix with lower dimensional ones with the same quantum numbers, with mixing coefficients which diverge as inverse powers of the lattice spacing. Thus for example, the kinetic energy operator in the HQET, $\bar{h}\vec{D}^2h$ (where $h$ represents the field off the heavy quark) mixes with $1/a^2\,\bar{h}h$ and $1/a\,\bar{h}D_4h$. These power divergences must be subtracted non-perturbatively in order to define "physical" matrix elements, since factors of the form

$$\exp\left[-\int^{g_0(a)}\frac{dg'}{\beta(g')}\right]=a\Lambda_{\rm QCD}\ ,\qquad(1)$$

combined with inverse powers of the lattice spacing can give finite contributions [13]. Renormalons are an example of such non-perturbative effects. Renormalons are also present in calculations using dimensional regularization for those quantities for which power divergences would be expected by power counting. Using dimensional regularization however, power divergences and the perturbative mixing of operators of different dimensions are absent because of the lack of a hard cut-off.

For most of this lecture I will consider computations in the HQET, however the principal features concerning the appearance and cancellation of renormalons are more general. Thus the discussion below can readily be generalized to other applications of operator product expansions. In the HQET one computes physical quantities as series in inverse powers of the heavy quark mass. The Dirac term in the QCD action is replaced by

$$\bar{Q}(i\slashed{D}-m_Q)Q\to\bar{h}_v(iv\cdot D)h_v+\frac{1}{2m_Q}\bar{h}_v(iD)^2h_v$$
$$+\frac{c_{\rm mag}}{2m_Q}\frac{g}{2}\bar{h}_v\sigma_{\alpha\beta}F^{\alpha\beta}h_v+O(1/m_Q^2)\ ,\qquad(2)$$

where $Q$ and $h_v$ represent the fields of the heavy quark in QCD and the HQET respectively, and $v$ is the quark's four-velocity. In all the simulations described below we will take the quark to be at rest, and will denote the corresponding heavy quark field by $h$ (without any subscript). $c_{\rm mag}$ is a constant determined by matching the effective theory onto the full one (QCD). The corresponding constant for the kinetic term

$(\bar{h}_v(iD)^2h_v)$ is equal to one by reparametrization invariance [14]. In addition to the expansion of the action in eq.(2), one must also expand the operators whose matrix elements are being computed (this will be discussed below). For a comprehensive review of the structure and applications of the HQET, and references to the original literature, see ref.[15], and for discussions of leading order computations using lattice simulations of the HQET see refs.[16,17].

Much of the discussion in this lecture will be based on the two interesting papers which pointed out the presence of renormalons in the HQET, and stressed that the pole-mass of a quark cannot be a physical parameter and should not be used as the expansion parameter [2,3]. The applications to the lattice formulation of the HQET, and the numerical results presented below, are based on work carried out with Guido Martinelli, Marco Crisafulli and Vicente Giménez [18,19].

The plan of this lecture is as follows. In the next section I present the main results concerning the appearance of renormalons in operator product expansions and their significance, delaying the corresponding detailed explanations until the later sections. In sections 3 and 4 I discuss the appearance of renormalons in Borel transforms of perturbation series, and present calculations of their effects in the limit of a large number of light quark flavours. The presence of renormalons in the pole mass in demonstrated in section 5. Section 6 contains a discussion of renormalons in the HQET regulated with a hard ultra-violet cutoff, and in the lattice formulation of the HQET in particular. Although the pole mass is not a physical parameter, short-distance definitions of the quark mass are physical, and in section 7 I explain how these can be evaluated from simulations in the HQET, and present some results for the $\overline{\rm MS}$ mass. The evaluation of the kinetic energy of the heavy quark is explained in section 8. This calculation demonstrates the general non-perturbative techniques needed to subtract power divergences and renormalon uncertainties from matrix elements of higher dimensional operators which were proposed in refs.[18,19]. In section 9 I discuss a possible definition of the parameter $\bar{\Lambda}$, the difference of the hadron and heavy quark masses, which is commonly introduced in applications of the HQET. Numerical results for $\bar{\Lambda}$ are also presented. Finally section 10 contains a summary and my conclusions.

## 2. Overview

In this section I will present the principal results concerning the appearance and cancellation of renormalon singularities in operator product expansions, using the Heavy Quark Expansion as an illustration. Consider the evaluation of the matrix element of some local QCD operator containing one or more heavy quark fields. We call this operator $O^{\mathrm{QCD}}$. For example if we wish to evaluate the leptonic decay constant $f_B$, then $O^{\mathrm{QCD}}$ would be the axial current $\bar{b}\gamma^\mu\gamma^5 q$, where $q$ represents the field of the light quark. Using the HQET we expand $O^{\mathrm{QCD}}$ in inverse powers of the mass of the heavy quark $m_Q$

$$O^{\mathrm{QCD}} = C_1(m_Q/\mu)\, O_1^{\mathrm{HQET}}(\mu) +$$
$$\frac{1}{m_Q}C_2(m_Q/\mu)\, O_2^{\mathrm{HQET}}(\mu) + O(1/m_Q^2) \qquad (3)$$

where $\mu$ is the scale at which the operators of the HQET, $O_1^{\mathrm{HQET}}$ and $O_2^{\mathrm{HQET}}$, are renormalized. We consider here the simple situation for which there is a single operator in each of the first two orders of the expansion, but the discussion below can be easily generalized to the case in which there are more operators.

When evaluating matrix elements of $O^{\mathrm{QCD}}$ beyond the leading order in the $1/m_Q$ expansion, in addition to the corresponding higher dimension operators on the r.h.s. of eq.(3), it is also necessary to keep higher order terms in the heavy quark action, as in eq.(2). Throughout the discussion below it is implied that this is done.

In general the QCD operator $O^{\mathrm{QCD}}$ also requires renormalization and is defined at some scale $M$. Unless specifically needed I will suppress the dependence on $M$ in $O^{\mathrm{QCD}}$ and in the coefficient functions $C_i$.

I will now summarize the main points which I wish to make in this lecture, postponing a detailed discussion to the following sections:

- If we restrict the calculation to the leading order in $1/m_Q$, i.e. if we neglect the $O(1/m_Q)$ terms, then the perturbation series for the coefficient function $C_1$ diverges, and is not Borel summable. As will be explained below this is due to a singularity in the Borel transform of the perturbation series, called an *infra-red renormalon*. Thus there is an ambiguity in the evaluation of $C_1$, coming from the different possible ways of defining the series. This ambiguity is of $O(1/m_Q)$.

- Formally therefore, we should not include the $O(1/m_Q)$ corrections in eq.(3) until we have computed sufficiently many terms in the perturbation series for $C_1$ to control its divergent behaviour.

- For this talk I will restrict the discussion to problems for which the matrix elements of $O^{\mathrm{QCD}}$ have no renormalon ambiguities. In general such non-perturbative effects do exist, and appear on the right hand side of eq.(3) in the matrix elements of the operators of the HQET. They do not affect the coefficient functions.

- In renormalization schemes based on the dimensional regularization of ultra-violet divergences in the HQET, such as the $\overline{\mathrm{MS}}$ renormalization scheme, the matrix elements of $O_2^{\mathrm{HQET}}$ are also not Borel summable in perturbation theory, due to an *ultra-violet renormalon* singularity in their Borel transforms. The corresponding ambiguities in the matrix elements of $O_2^{\mathrm{HQET}}$ cancel those in $C_1$.

- If a hard ultra-violet cut-off is used, such as the lattice spacing in the lattice HQET, then the matrix elements of $O_2^{\mathrm{HQET}}$ do not have ambiguities due to ultra-violet renormalons. This, in turn, implies that the coefficient functions $C_1$ do not contain ambiguities due to infra-red renormalons. In the lattice theory it is natural to present the discussion in terms of bare operators in the effective theory, defined with the lattice spacing $a$ as the ultra-violet cut-off. Throughout this talk I will assume that $m_Q a \gg 1$.

Of course if the inverse lattice spacing was much smaller than the heavy quark mass then there would be no need to use the HQET.

- The absence of renormalon ambiguities in $C_1$ in the lattice theory is due to a cancellation between leading and non-leading terms in perturbation theory, i.e. terms which behave logarithmically with $m_Q a$ and non-leading ones of $O(1/m_Q a)$.

- In the lattice theory, matrix elements of higher dimensional operators (such as $O_2^{\mathrm{HQET}}$) diverge as inverse powers of the lattice spacing, and are hence manifestly unphysical. For example

$$\langle H|\bar{h}\vec{D}^2 h|H\rangle \sim O(1/a^2) \qquad (4)$$

where $H$ represents the heavy hadron.

- The subtraction of the power divergences in perturbation theory introduces renormalon ambiguities. Thus for example, the perturbation series of the terms which diverge quadratically, i.e. those which are $O(1/a^2)$, in eq.(4) contains a renormalon ambiguity.

- The perturbation series for the pole mass has a renormalon ambiguity of $O(\Lambda_{\mathrm{QCD}})$ [2, 3]. This is not the case for short-distance definitions of the heavy quark mass, such as

$$\overline{m}_Q \equiv m_Q^{\overline{\mathrm{MS}}}(m_Q^{\overline{\mathrm{MS}}}) . \qquad (5)$$

- It is possible to compute $\overline{m}_Q$ (and other short-distance definitions of the mass), using only simulations in the HQET. In ref.[19] it was found that

$$\overline{m}_b = 4.17 \pm 0.05 \pm 0.03 \,\mathrm{GeV} + O(1/m_b) . (6)$$

- The presence of renormalons in physical quantities for which there is no Operator Product Expansion, such as the Drell-Yan process or in event shape variables in jet physics, is a subject currently under intensive investigation [7]-[12]. It is hoped that these studies will provide important phenomenological information about the sub-asymptotic (non-leading twist) behaviour of physical quantities. In lattice QCD there are analogous questions about the presence of ambiguities in perturbation series for quantities which contain power divergences, e.g. is there an ambiguity in the perturbative evaluation of the critical mass when using Wilson fermions? I will return to these questions at the end of this talk.

## 3. The Borel Transform and Renormalons

In this section I define the Borel transform of a perturbation series and explain what is meant by a renormalon singularity. Consider the perturbation series for some quantity $F$:

$$F(\alpha_s) = \sum_{n=0}^{\infty} F_n \left( \frac{\beta_0}{4\pi} \alpha_s(\mu) \right)^n \qquad (7)$$

where, for later convenience, we choose to include $\beta_0/4\pi$ in the expansion parameter ($\beta_0 = 11 - 2/3\, n_f$ is the first term in the perturbative expansion of the $\beta$-function). The Borel transform $\tilde{F}(u)$ of $F(\alpha_s)$ is defined by

$$\tilde{F}(u) = F_0 \delta(u) + \sum_{n=0}^{\infty} \frac{1}{n!} F_{n+1} u^n . \qquad (8)$$

If the theory is Borel summable then F is reconstructed by taking the Laplace transform

$$F(\alpha_s) = \int_0^{\infty} du \exp\left( \frac{-4\pi u}{\beta_0 \alpha_s(\mu)} \right) \tilde{F}(u) . \qquad (9)$$

$F(\alpha_s)$ is the unique function which satisfies certain analyticity conditions in the $\alpha_s$-plane with the asymptotic expansion (7) (see ref.[20] for example).

If the coefficients $F_n$ grow like $n!$ or faster, then the factor of $n!$ in the denominator of the terms on the r.h.s. of eq.(8) will not be sufficient to prevent singularities appearing in $\tilde{F}(u)$ for values of $u$ on the positive real axis. In such cases the integral in eq.(9) is not defined. Of course the contour of integration in eq.(9) can be deformed away from the singularities, but in general the resulting integral will depend on how this is done, leading to an ambiguity in the result. *Renormalons* is the name given to the singularities on the positive real axis in the $u$-plane.

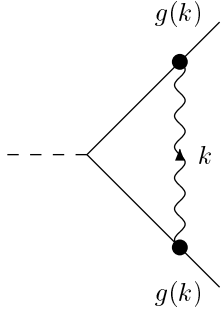

Figure 1. One loop graph contributing to the total cross section for $e^+e^-$ annihilation. The dashed line repesents a virtual photon or $Z$-Boson.

In order to obtain some intuition concerning the appearance of renormalons consider the one loop graph in Fig.1, which can be considered, for example, as a contribution to the amplitude for $e^+e^- \rightarrow$ hadrons. It may be expected that part of the effect of higher order corrections will be to renormalize the coupling constant, and so we take the coupling constants at the vertices to be $g(k)$, where $k$ is the momentum in the gluon propagator. Now expanding $g(k)$ as a power series in $g(\mu)$, where $\mu$ is some fixed renormalization scale, we readily find that the coefficients in perturbation series grow like $n!$.

In general the renormalon singularities are branch points of cut singularities, whose position and nature are determined by the renormalization group, but not their residues. In some simplified examples below, motivated by the large $N_f$ expansion, where $N_f$ is the number of quark flavours, they appear as simple poles:

$$\tilde{F}(u) = \sum_i \frac{r_i}{u - u_i} + \cdots \qquad (10)$$

where we choose the ordering $0 < u_1 < u_2 < u_3 \cdots$. Writing

$$\frac{1}{u - u_i} = \frac{1}{(u - u_i)_\mathcal{P}} + \eta_i \, \delta(u - u_i) \qquad (11)$$

where the subscript $\mathcal{P}$ stands for "principal part", and $\eta_i$ defines the prescription used to define the pole. We define $\Delta F$, the ambiguity in $F$ as the coefficient of $\eta_1$ [5],

$$\Delta F = r_1 \exp\left(\frac{-4\pi u_1}{\beta_0 \alpha_s(\mu)}\right) = r_1 \left(\frac{\Lambda_{\text{QCD}}}{\mu}\right)^{2u_1}. \quad (12)$$

From eq.(12) we see that the ambiguity is indeed "exponentially small", and hence is suppressed by a power of $\Lambda_{\text{QCD}}/\mu$.

## 4. The Large $N_f$ Limit

Of course in general it is not possible to sum all higher order graphs completely. Much insight can be obtained by studying the behaviour of perturbation series in the limit of a large number of light quark flavours $N_f$. In this limit the gluon propagator is given by the sum of bubble diagrams shown in Fig.2, since it is only for these graphs that each additional factor of $\alpha_s$ is accompanied by a factor of $N_f$. The geometric series for the gluon propagator in this limit can readily be summed, leading to the following expression for its Borel Transform in the Landau gauge:

$$\tilde{D}_{ab}^{\mu\nu}(k, u) = i\delta_{ab} \left(\frac{\text{e}^C}{\mu^2}\right)^{-u} \frac{k^\mu k^\nu - k^2 g^{\mu\nu}}{(-k^2)^{2+u}}, \quad (13)$$

where $a$ and $b$ are colour labels. $C$ is the finite (renormalization-scheme dependent) term in the one-loop graph, which is given by

$$-\frac{\alpha_s N_f}{6\pi} \left(\ln(-k^2/\mu^2) + C\right) . \qquad (14)$$

In the large $N_f$ limit asymptotic freedom is lost, and instead we replace the factor $-2/3N_f$ in eq.(14) by $\beta_0$. In this way we sum the leading powers of $\beta_0$ in each order of perturbation theory, and the procedure is motivated by the intuition that it is the scale dependence of the coupling constant which gives rise to renormalons, at least partially.

In the approximation described above, all the dependence on the coupling constant comes from diagrams with a single resummed gluon propagator. Thus the $O(1/N_f)$ contribution to the Borel transform of the quark propagator is given by the diagram in Fig.3 (this is the leading non-trivial contribution). In the following section we study the graph in Fig.3, and demonstrate the appearance of renormalons in the Borel transform of the "pole-mass".

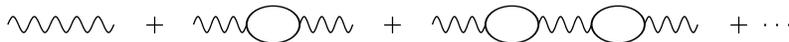

Figure 2. The sum of diagrams contributing to the gluon propagator in the large $N_f$ limit. The "bubbles" represent loops of light quarks.

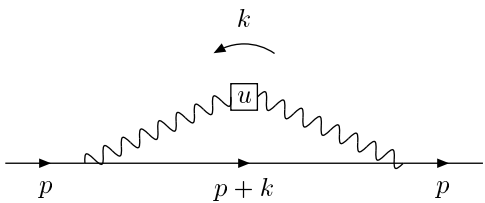

Figure 3. Diagram contributing to the leading non-trivial term in the Borel transform of the quark propagator in QCD. The box containing $u$ implies that for the gluon propagator we take the expression in eq.(13).

## 5. Renormalons in the Pole Mass

We are now ready to study the Borel transform of the perturbation series for the pole mass. The pole mass is given by

$$m_{\text{pole}} = m_Q + \Sigma(\displaystyle{\not}p)_{\displaystyle{\not}p = m_{\text{pole}}} \tag{15}$$

where $m_Q$ is the bare heavy-quark mass and $\Sigma$ is the self-energy diagram, whose Borel transform is represented in Fig.3. Eq.(15) is an implicit equation for the pole mass, however, as

$$m_{\text{pole}} = m_Q + O\left(\frac{1}{N_f}\right) , \tag{16}$$

in the large $N_f$ limit we can replace $m_{\text{pole}}$ with $m_Q$ on the right hand side of eq.(15). Thus the Borel transform of the pole mass can be obtained by simply evaluating the graph in Fig.3, on shell, i.e. with $p^2 = m_Q^2$.

It is straightforward to see that on-shell there is a pole singularity at $u = 1/2$. The infra-red behaviour of the graph in Fig.3 is given by

$$\int_{k \to 0} d^4k \frac{1}{(k^2)^{1+u}k} \tag{17}$$

which, using dimensional regularization, has pole singularities of the form $\Gamma(1-2u)$, i.e. the first i.r. renormalon appears at $u = 1/2$. The corresponding ambiguity in the pole mass is of $O(\Lambda_{\text{QCD}})$ (see eq.(12) above):

$$\Delta m_{\text{pole}} = -\frac{2C_F}{\beta_0}e^{-C/2}\Lambda_{\text{QCD}} . \tag{18}$$

For the following discussion it will be useful to observe that off-shell, i.e. for $p^2 \neq m_Q^2$ there is no pole at $u = 1/2$. In this case the infra-red behaviour of the graph in Fig.3 is given by

$$\int_{k \to 0} d^4k \frac{1}{(k^2)^{1+u}} \tag{19}$$

which is manifestly finite for $u = 1/2$.

We are now ready to demonstrate the cancellation of renormalons described in section 2. Consider the expansion of the inverse heavy quark propagator in QCD, as a series in inverse powers of the heavy-quark mass. In perturbation theory it takes the form:

$$S_P^{-1}(p) = m - m_{\text{pole}} + C(m_Q/\mu)S_{\text{eff}}^{-1}(v \cdot k, \mu)$$
$$+ O(1/m_Q) \tag{20}$$

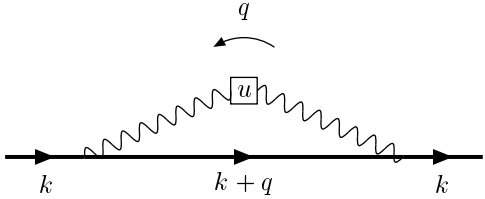

Figure 4. Diagram contributing to the leading non-trivial order term in the Borel transform of the quark propagator in the HQET (the bold line represents the quark propagator in the effective theory). The box containing $u$ implies that for the gluon propagator we take the expression in eq.(13).

where $S_{\text{eff}}^{-1}$ is the inverse of the heavy quark propagator in the HQET. The subscript $P$ implies the projection:

$$S_P^{-1} \equiv \frac{1 + \not{v}}{2} S^{-1} \frac{1 + \not{v}}{2} \,. \tag{21}$$

$k$ is defined by $p = mv + k$, where $m$ will be chosen conveniently later. The freedom to choose $m$ in this decomposition of $p$ is equivalent to the freedom to choose a renormalization condition in the HQET to fix the mass counterterm.

The Borel transform of eq.(20) takes the form:

$$\tilde{S}_P^{-1}(p, u) = \tilde{m}(u) - \tilde{m}_{\text{pole}}(u) +$$
$$\tilde{C}(m_Q/\mu, u) * \tilde{S}_{\text{eff}}^{-1}(v \cdot k, \mu, u) + O(1/m_Q) \tag{22}$$

where the tilde represents the Borel transform and $*$ the convolution. We have seen above that $\tilde{m}_{\text{pole}}$ has an i.r. renormalon at $u = 1/2$. $\tilde{S}_{\text{eff}}^{-1}$ has an u.v. renormalon at u=1/2. This can easily be deduced by power counting. The graph in Fig.4 has the ultra-violet behaviour

$$\int_{\text{large } q} d^4q \, \frac{1}{(q^2)^{1+u} q} \tag{23}$$

which yields a pole at $u = 1/2$.

The i.r. renormalon in $\tilde{m}_{\text{pole}}$ is cancelled by the u.v. renormalon in $\tilde{S}_{\text{eff}}^{-1}$. This can be verified by explicit calculation, but follows directly from the observation made above, that the Borel transform of $S^{-1}(p)$ has no renormalon at $u = 1/2$ for off-shell values of the momenta $p$.

The connection between the presence of power divergences in perturbation theory and renormalons can also be demonstrated using power counting. The integral in eq.(23) has a logarithmic ultra-violet divergence at $u = 1/2$, which implies that it has a power (in this case linear) divergence at $u = 0$. The one-loop contribution to the quark self energy corresponds to $u = 0$. Conversely, if a low order graph with a single gluon propagator has a power divergence in perturbation theory, then the Borel Transform of the corresponding series of graphs with bubble insertions on the gluon propagator has singularities at positive values of $u$, i.e. renormalon singularities.

The appearance and cancellation of renormalons in eq.(22) is an example of the general structure outlined in section 2. In this case we can view $O^{\text{QCD}}$ as the corresponding Dirac operator in QCD, and $O_1^{\text{HQET}}$ and $O_2^{\text{HQET}}$ as $\bar{h}_v h_v$ and $\bar{h}_v v \cdot D h_v$ respectively.

## 6. Renormalons and Lattice Perturbation Theory

The discussion in the previous section was based on the use of dimensional regularization of the ultra-violet divergences. With a hard ultra-violet cut-off, such as the lattice spacing, there cannot be a renormalon singularity in $S_{\text{eff}}^{-1}$ (see eq.(20)). This is because the cut-off itself ensures that the integral corresponding to eq.(23) is finite at all values of $u$.

In lattice perturbation theory, the inverse propagator in the HQET takes the form

$$S_{\text{eff}}^{-1}(v \cdot k) = v \cdot k \, [1 + \frac{\alpha_s(a) C_F}{4\pi} \left(-\gamma_h \ln(-2av \cdot k)\right.$$
$$\left. + d\right)] - \alpha_s(a) \frac{X}{a} + \cdots \,, \tag{24}$$

where $\gamma_h$ is the one-loop contribution to the anomalous dimension of the heavy quark field, $\alpha_s(a)$ is the bare coupling constant and $d$ and $X$ are constants. The presence of the linearly divergent mass term, $-\alpha_s(a)\frac{X}{a}$, is a manifestation of the mixing of the operator $\bar{h}_v v \cdot Dh_v$ with the lower dimensional one $\bar{h}_v h_v$.

As mentioned above the Borel transform of the series in eq.(24) does not have any renormalon singularities. However the absence of these singularities occurs through an interesting cancellation. The series of terms in eq.(24) which diverge linearly does have a renormalon at $u = 1/2$. This renormalon cancels against that present in the Borel transform of the series of terms proportional to $v \cdot k$. To see this consider the graph in Fig.4 for $v \cdot k = 0$. The infra-red behaviour of the graph is given by

$$\int_{q\to 0} d^4 q \frac{1}{(q^2)^{1+u} q} \ , \qquad (25)$$

which has pole at $u = 1/2$, demonstrating the existence of the renormalon in the series of terms which diverge linearly in each order of perturbation theory. Since at $v \cdot k \neq 0$ there is no singularity, the renormalon is cancelled by that in the terms which, in each order, are proportional to $v \cdot k$. The structure of the Borel transform of the inverse propagator as $u \to 1/2$ is:

$$\tilde{S}_{\text{eff}}^{-1}(v \cdot k, u) \sim \frac{1}{1-2u} \left[ v \cdot k (av \cdot k)^{-2u} - \frac{1}{a} \right] \ (26)$$

plus terms which are manifestly non-singular at $u = 1/2$. The residue of the pole at $u = 1/2$ is zero as a result of the cancellation described above.

## 7. The $\overline{\text{MS}}$ Mass

Although the pole mass is ambiguous due to the presence of renormalon singularities, it is possible to evaluate any "short-distance" mass using lattice simulations in the HQET and perturbation theory. In this section we present a determination of the $\overline{\text{MS}}$ mass $\overline{m}_Q$ defined in eq.(5). Consider the correlation function of some heavy-light hadron, for example that of the pseudoscalar meson $P$,

$$C(t) = \sum_{\vec{x}} \langle 0 | \bar{h}(x) \gamma^5 q(x) \bar{q}(0) \gamma^5 h(0) | 0 \rangle \qquad (27)$$

$$= Z^2 \exp(-\epsilon t) \ . \qquad (28)$$

The matrix element $Z$ is proportional to the decay constant $f_P$ in the static approximation, and has been the subject of many lattice computations [17]. The quantity $\epsilon$ is linearly divergent as $a \to 0$. Now $\overline{m}_Q$ can be obtained from the computed value of $\epsilon$ and the known value of the mass of the meson $P$ (in practice that of the $B$-meson) by the perturbative relation [19]

$$\overline{m}_Q = \left( m_P - \epsilon + \frac{\alpha_s X}{a} + \cdots \right)$$
$$\times \left( 1 - \frac{4\alpha_s}{3\pi} + \cdots \right) \ . \qquad (29)$$

The perturbation series in each of the two factors in eq.(29) have renormalon singularities at $u = 1/2$, and these singularities cancel in the product. The first factor corresponds, in perturbation theory, to the determination of $m_{\text{pole}}$ from the lattice simulation of the effective theory. The second factor is the perturbation series relating $m_{\text{pole}}$ to $\overline{m}_Q$. In this way $\overline{m}_Q$ can be determined from the computation of $\epsilon$. The remaining ambiguities in eq.(29) correspond to renormalons at $u = 1$, and hence are of $O(\Lambda_{\text{QCD}}^2/m_Q)$. Since in the present discussion we do not include $1/m_Q$ corrections to the HQET action, these remaining ambiguities are of the same order as other uncertainties. In principle we can also systematically reduce the ambiguities and uncertainties to higher orders in $1/m_Q$.

We have recently determined $\overline{m}_b$ in this way, using the precise results for $\epsilon$ obtained by the APE collaboration and found [19]:

$$\overline{m}_b = 4.22(7) \text{ GeV} \ \text{at} \ \beta = 6.0 \qquad (30)$$
$$\overline{m}_b = 4.15(8) \text{ GeV} \ \text{at} \ \beta = 6.2 \ . \qquad (31)$$

Combining the results in eqs.(30) and (31) we obtain our best estimate which has been given in eq.(6) above.

In obtaining these results we have used one-loop perturbation theory in the matching equation (29). It will be very important to extend this to higher orders to be confident that we control the cancellation of the renormalon singularities with sufficient precision.

## 8. The Kinetic Energy Operator

In this section I discuss the evaluation of the matrix elements of the kinetic energy operator which is proportional to $\bar{h}\vec{D}^2h$. These matrix elements are needed, for example, in phenomenological applications of the HQET to the spectroscopy and inclusive decays of heavy hadrons. This calculation provided an illustration of the general techniques we propose for the subtraction of power divergences, which appear due to the mixing of higher dimensional operators with lower dimensional ones with the same quantum numbers.

We start with the observation that the matrix elements of the bare lattice operator are huge, e.g. from a simulation at $\beta = 6.0$ on a $16^3 \times 32$ lattice using the SW-Clover quark action at $\kappa = 0.1425$ we found [19],

$$a^2 \frac{\langle B|\bar{h}\vec{D}^2h|B\rangle}{2M_B} = -0.72 \pm 0.14 \ , \tag{32}$$

which corresponds in physical units to

$$\frac{\langle B|\bar{h}\vec{D}^2h|B\rangle}{2M_B} \simeq -2.9 \ \text{GeV}^2 \tag{33}$$

whereas for a physical matrix element we would expect the result to be of $O(\Lambda_{\text{QCD}}^2)$. Of course the matrix element of the bare operator is not a physical quantity, indeed, as explained above, it is quadratically divergent as $a \to 0$. However the result above illustrates that the necessary subtractions will be very large, and hence the determination of physical effects will be difficult. In one-loop perturbation theory the term to be subtracted is $-5.19 \, \alpha_s/a^2$ which is of the right order of magnitude. However, the series of quadratically (and linearly) divergent terms have renormalon ambiguities and hence a subtraction of these divergences in perturbation theory does not lead to a physical matrix element. It is therefore necessary to perform these subtractions nonperturbatively [13]. We propose to do this by imposing a suitable renormalization condition on the matrix element taken between heavy quark states, and to compute the corresponding subtraction constant using lattice simulations. For example we can impose the natural condition that

$$\langle h(\vec{p}=0)|\bar{h}\vec{D}_S^2h|h(\vec{p}=0)\rangle = 0 \tag{34}$$

in some gauge, e.g. the Landau gauge, where the subtracted operator $\bar{h}\vec{D}_S^2h$ is given by a linear combination of $\bar{h}\vec{D}^2h$ and all lower dimensional ones with which it can mix

$$\bar{h}\vec{D}_S^2h = \bar{h}\vec{D}^2h - c_1\bar{h}\vec{v}\cdot Dh - c_2\bar{h}h \ . \tag{35}$$

For the matrix elements of the subtracted operator between hadronic states it is the constant $c_2$ which is required (the term multiplying $c_1$ can be eliminated by the equations of motion), and this can be determined with very good precision [19]

$$a^2 c_2 = -0.73 \pm 0.01 \pm 0.02 \ . \tag{36}$$

The matrix element of the subtracted operator in lattice units is then given by the difference of the results in equations (32) and (36). The cancellation is overwhelming and leads to very large relative errors. Clearly a very large statistical sample will be required to obtain a result for the matrix element of the subtracted kinetic energy operator.

## 9. $\bar{\Lambda}$

Much of the phenomenology using the HQET is presented in terms of the parameter $\bar{\Lambda} = m_H - m_Q$, where $m_H$ is the mass of the heavy light hadron $H$. In view of the discussion above the question arises as to what we should take for the heavy quark mass $m_Q$. The pole mass is ambiguous and cannot be used [2,3]. Short-distance definitions of the heavy quark mass can be used safely, but then it is not clear whether the introduction of $\bar{\Lambda}$ is useful (e.g. in that case $\bar{\Lambda}$ does not go to a constant as $m_Q \to \infty$).

In section 5 we have seen that even if we start with the bare HQET with no mass term, $\mathcal{L} = \bar{h}D_4h$, the presence of power divergences and renormalons generates a mass term. It is therefore possible to add a mass term $(\delta m\bar{h}h)$ to the action and to fix the counterterm by imposing a renormalization condition on the heavy quark propagator. Lattice simulations provide the opportunity of determining the counterterm nonperturbatively. The procedure would be as follows:

- Consider the theory with no mass term, $\mathcal{L} = \bar{h} D_4 h$. Compute the heavy quark propagator in a fixed gauge (the Landau gauge for example), and determine the effective mass $m_{\text{eff}}(t)$; $a m_{\text{eff}}(t) = \ln(S_{\text{eff}}(t)/S_{\text{eff}}(t+a))$. Results for the effective mass from simulations at $\beta = 6.0, 6.2$ and $6.4$ are presented in Fig.5.

- Add a mass counterterm $\delta m \bar{h} h$ to the action. The propagator now changes by a factor $\exp(-\mu t)$, where $\mu = 1/a \ln(1 + \delta m\, a)$, and hence the effective mass at any time changes by $\mu$.

- Impose some renormalization condition on the effective mass, such as $m_{\text{eff}}(t^*) = 0$ for some fixed $t^*$. At small times we also have a perturbative contribution to the effective mass

$$m_{\text{eff}}(t) = -\frac{\alpha_S C_F}{4\pi} \frac{\gamma_h}{t} + \cdots \qquad (37)$$

where $\gamma_h$ is the one-loop contribution to the anomalous dimension of the heavy quark field (in the Landau gauge $\gamma_h = -6$). Such a behaviour is clearly seen in Fig.5.

- For large values of $t$ we are in the non-perturbative regime, and do not know the behaviour of the propagator. We know that there is an exponential factor from the renormalon singularity. If other non-perturbative effects do not change the effective mass at large values of $t$, then $m_{\text{eff}}$ will tend to a constant as $t \to \infty$. Lattice simulations provide the opportunity to investigate whether this is true. The results presented in Fig.5 are consistent with the assumption that the effective mass tends to a constant at large times. If this is the case then a natural definition of the mass counterterm is to impose that the effective mass is zero at large times, and define $\bar{\Lambda}$ by

$$\bar{\Lambda} = \epsilon + \mu \qquad (38)$$

and a subtracted pole mass $m_Q^S$ by

$$m_Q^S = m_H - \bar{\Lambda}\ , \qquad (39)$$

where $H$ is the heavy hadron [18]. $\bar{\Lambda}$ and $m_Q^S$ defined in this way are free of power divergences and renormalon ambiguities, and are independent of the method used to regulate the ultra-violet divergences. It must be stressed however, that this definition of $\bar{\Lambda}$ assumes properties of the long-distance behaviour of the quark propagator.

A further check on this procedure from lattice simulations is that the result for $\bar{\Lambda}$ should be independent of the lattice spacing. Within our errors this appears to be the case. We find [19]

$$\bar{\Lambda} = 180 \pm 35 \text{ MeV} \text{ at } \beta = 6.0 \qquad (40)$$

$$\bar{\Lambda} = 220 \pm 55 \text{ MeV} \text{ at } \beta = 6.2 \qquad (41)$$

where the errors combine statistical uncertainties with those due to the calibration of the lattice spacing. Combining the results we quote

$$\bar{\Lambda} = 190 \begin{array}{c} + 50 \\ - 30 \end{array} \text{ MeV} \qquad (42)$$

as our best estimate of $\bar{\Lambda}$.

I end by emphasising the obvious but important point, that when presenting a value for $\bar{\Lambda}$, one must be careful in defining precisely what is meant by this parameter. In particular it is not sufficient to say that the quark mass is the pole mass.

## 10. Conclusions

In this talk I have attempted to summarize the considerable progress which has been made recently towards the theoretical understanding of higher twist (or dimension) corrections to physical processes. It can now be expected that a sound phenomenology of these corrections can be developed. A central step in this is the complete definition of higher dimensional operators (we have seen that it is not sufficient to say that they are defined in the $\overline{\text{MS}}$ scheme). The most important points have been summarized in the overview presented in section 2, which should be considered as the main part of the conclusions.

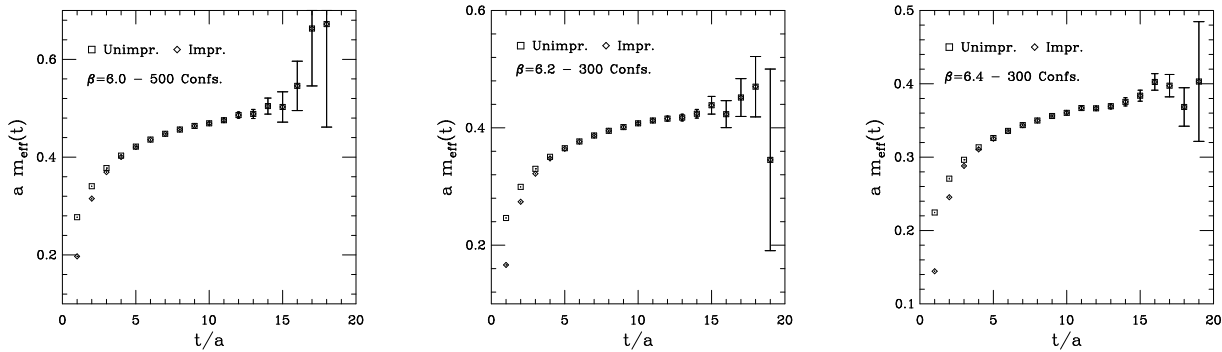

Figure 5. The Heavy Quark Effective Masses from simulations by the APE collaboration at 3 different values of $\beta$. In each case the size of the lattice is $24^3 \times 40$.

I conclude with a brief mention of some related questions. In refs.[7–9] the question of sub-asymptotic corrections to the Drell-Yan process is studied. In refs.[7,8] it is argued that the presence of the leading i.r. renormalon implies that the leading power corrections are of $O(1/Q)$, where $Q$ is the invariant mass of the lepton pair. In ref.[9] it is shown that the coefficient of this renormalon is zero in the large $N_f$ limit. It is a very important question whether this is true beyond this approximation. In any case, from studies of the renormalons, it is agreed that there are $1/Q$ corrections to event-shape variable in $e^+e^-$ collisions.

It is an interesting question whether there are renormalon ambiguities in lattice perturbation theory for quantities which diverge like inverse powers of the lattice spacing. A simple example is the critical mass, $m_{\rm crit}$ (for which the mass of the pion is zero) in simulations of lattice QCD with Wilson fermions. In the large $N_f$ limit the i.r. renormalon at $u = 1/2$ is absent, implying that there is no ambiguity in $m_{\rm crit}$ of $O(\Lambda_{\rm QCD})$ [21]. It will be interesting to establish whether this result survives in higher orders in the $N_f$ expansion.

In the large $N_f$ limit the coefficient of the i.r. renormalon at $u = 1$ in the pole mass vanishes [3], and equivalently the u.v. renormalon at $u = 1$ in the kinetic energy operator also has zero coefficient. In ref.[22] this was called the *invisible renormalon*. So far it is not understood whether there is some symmetry which would allow one to conclude that this is a general result rather than an accident of the large $N_f$ limit. In the lattice formulation of the HQET the corresponding quadratic divergence is present [13], whereas using Pauli-Villars or momentum-flow regularization the coefficient of the quadratic divergence at one-loop order is zero [22].

We have seen that the cancellation of renormalons in Wilson coefficient functions occurs between terms which in finite orders of perturbation theory are of different orders in the expansion in $m_Q$ or other asymptotic variable. The higher dimension operators must be defined completely. Ji has revisited the extraction of the gluon condensate from the lattice measurements of the plaquette $\mathcal{P}$ [23]

$$1 - \frac{1}{N_c}\mathrm{Tr}\mathcal{P} = \sum_n \frac{c_n}{\beta^n} + \\ \frac{\pi^2}{12N_c}a^4(\beta)\langle\frac{\alpha_s}{\pi}F^2\rangle + O(a^6) \ . \qquad (43)$$

Clearly before we can determine the gluon condensate from the lattice measurement we have to subtract the perturbative series. This has to be done carefully as the series has i.r. renormalon singularities. It is remarkable that, using the Langevin stochastic formulation of the lattice Yang-Mills theory, the Parma group were able to determine the perturbative coefficients up to eight-loop order [24]!

$$\sum_n \frac{c_n}{\beta^n} = \frac{2}{\beta} + \frac{1.218}{\beta^2} + \frac{2.960}{\beta^3} +$$

$$\frac{9.28}{\beta^4} + \frac{34}{\beta^5} + \frac{135}{\beta^6} + \frac{563}{\beta^7} + \frac{2488}{\beta^8} + \cdots . \ (44)$$

It seems likely that these techniques will be very important in other applications where high orders of perturbation theory are needed to control the cancellation of renormalon ambiguities. Ji's attempts to define a gluon condensate by subtracting various resummed versions of the series in eq.(44) led to results which were a factor of more than 5 larger than values used in QCD sum rule applications [23]. It remains an interesting question, from both the theoretical and numerical point of view, how to connect lattice determinations of the condensates with those used in sum rules? Ref.[25] contains some discussion concerning the definition of condensates in sum-rule applications.

**Acknowledgements:** I warmly thank my collaborators, Marco Crisafulli, Vicente Giménez, Guido Martinelli, Matthias Neubert and Juan Nieves with whom my understanding of many of the issues discussed in the lecture has developed. I am also grateful to Martin Beneke, Volodya Braun, Al Mueller and Arkady Vaishtein for many helpful and stimulating discussions. I acknowledge the support of the Particle Physics and Astronomy Research Council, UK through the award of a Senior Fellowship.